\documentstyle{mn}
\input epsf
\def\spose#1{\hbox to 0pt{#1\hss}}
\newcommand{\approxlt}{\mathrel{\spose{\lower 3pt\hbox{$\sim$}}
	\raise 2.0pt\hbox{$<$}}}
\newcommand{\approxgt}{\mathrel{\spose{\lower 3pt\hbox{$\sim$}}
	\raise 2.0pt\hbox{$>$}}}
\newcommand{\ergpcmsqps}{{\rm erg}\,{\rm cm}^{-2}\,{\rm s}^{-1}\,}

\title[The colour of the Deep $ROSAT$ X-ray sky fluctuations]
{The colour of the Deep $ROSAT$ X-ray sky fluctuations}

\author[M.T. Ceballos, X. Barcons \& F.J. Carrera]
	{ M.T. Ceballos$^{1,2}$, X. Barcons$^{1}$ and F.J. Carrera$^{1}$
	\\
	$^1$ Instituto de F\'{\i}sica de Cantabria (Consejo Superior
	de Investigaciones Cient\'{\i}ficas - Universidad\\ 
	de Cantabria), 39005 Santander, Spain\\
	$^2$ Departamento de F\'{\i}sica Moderna, Universidad de
	Cantabria, 39005 Santander, Spain}

\begin{document}

\maketitle

\begin{abstract}      
 
We have carried out a fluctuation analysis in four bands (R4, R5, R6
and R7 corresponding to 0.44-1.01~keV, 0.56-1.21~keV, 0.73-1.56~keV
and to 1.05-2.04~keV respectively) of two very deep $ROSAT$ PSPC
observations in directions where the galactic H{\small I} column is
minimal. This has enabled us to study the average spectrum of the
sources contributing to 0.5-2~keV fluxes $\sim 10^{-15} \rm
\ergpcmsqps$. The best fit spectral energy index for the average faint
source spectrum, $\alpha=0.95^{+0.10}_{-0.15}$, is still steeper than
the one measured for the extragalactic XRB at these energies
$\alpha\sim 0.4-0.7$, but flatter than the typical AGN spectral index
$\alpha\sim 1-1.5$. This result has allowed us to constrain the
existence of a population of sources harder than the AGNs contributing
to the source counts at these fluxes. We find that a population of
X-ray sources with energy spectral index $\alpha\sim 0.4-0.6$ would
contribute $(30\pm 20)$ per cent to the source counts at these fluxes
if the rest of them are AGNs with energy spectral index $\alpha\sim
1.2$.

\end{abstract}


\section{INTRODUCTION}

The existence of a population, different to the AGNs, of faint sources
with hard X--ray spectra has been proposed in several
works. Hasinger et al. (1993) considered these sources as a possible explanation
for the discrepancy between the source number counts found in their
analysis and the number counts predicted by the integration of models
for the AGN X-ray luminosity function at fluxes $ \rm S\sim
2.5\,\times 10^{-15} \rm \ergpcmsqps$. The idea is also supported by
the tendency shown by the sources in their sample to be harder at
fainter fluxes (see also Vikhlinin et al. 1995 for further evidence on
this hardening).

Moreover, should these sources exist they could help to solve the
spectral paradox concerning the origin of the soft X--ray background:
the fact that while the spectral shape of the cosmic background at
energies $\approxgt 0.8$ keV is consistent with an energy spectral
index $\alpha\sim 0.4$ (Gendreau et al. 1995) or $\alpha\approxlt 0.6$
in the 0.5-2~keV band (Barber \& Warwick 1994, Chen et al. 1994), the
mean spectral index of the Broad Line AGNs in the 0.5-2~keV band,
which are thought to be its main contributors (Shanks et al. 1991,
Boyle et al. 1994, Hasinger et al. 1993), is $\approxgt 1.0$.

Recent work by Romero-Colmenero et al. (1996) and Griffiths et al.
(1996) has suggested that this new population might be dominated by
Narrow Emission Line Galaxies having a mean spectral slope of
$\alpha \sim 0.4$. The ratio between the number density of these
galaxies and the Broad Line AGNs is found to increase towards fainter
fluxes pointing to these sources as important contributors to
the extragalactic X--ray background at soft energies ($< 2\,\rm keV$).

\begin{table*}
  \vbox to 0cm{\vfil} 
\centering
\begin{minipage}{22cm} 
\caption{Data images from the Deep surveys}
 \label{imagenes} 
\begin{tabular}{ | l c c c c c| }
\hline Image & Channels & Exposure & Clean & Galactic H{\small I} & Mean \\ 
Name & & Time & Time & Column Density & counts \\
     & & (seconds)& (seconds)& ($\rm cm^{-2}$) & ($\rm counts\,arcmin^{-2}$)\\
\hline
 UKDS-R4 & 52 - 69 & 73311 & 69525 &$ 6.5\times10^{19}$ & $6.64\pm0.12$\\
 UKDS-R5 & 70 - 90 & 73311 & 69525 &$ 6.5\times 10^{19}$ & $4.91\pm0.12$\\
 UKDS-R6 & 91 - 131 & 73311 & 69525 &$ 6.5\times 10^{19}$ & $4.64\pm0.15$\\
 UKDS-R7 & 132 - 201 & 73311 & 69525 &$ 6.5\times 10^{19}$ & $2.88\pm0.10$ \\
 UKDS-HARD & 40 - 200 & 73311 & 69525 &$ 6.5\times 10^{19}$ & $25.19\pm0.41$ \\
\hline
 LHDS-R4 & 52 - 69 & 143700 & 93129 &$ 6.0\times 10^{19}$ & $11.48\pm0.13$ \\
 LHDS-R5 & 70 - 90 & 143700 & 93129 &$ 6.0\times 10^{19}$ & $6.73\pm0.14$ \\
 LHDS-R6 & 91 - 131 & 143700 & 93129 &$ 6.0\times 10^{19}$ & $5.98\pm0.16$ \\
 LHDS-R7 & 132 - 201 & 143700 & 93129 &$ 6.0\times 10^{19}$ & $3.91\pm0.12$ \\
 LHDS-HARD & 40 - 200 & 143700 & 93129 &$ 6.0\times 10^{19}$ & $36.09\pm0.42$\\
\hline
   \end{tabular}
 \end{minipage}
\end{table*}

In this work we have carried out an analysis of the sky fluctuations
in four spectral bands of two deep $ROSAT$ observations. This analysis
has allowed us to constrain the summed spectral shape of the sources
contributing at fluxes where this technique is sensitive ($\sim
10^{-15} \rm \ergpcmsqps$, 0.5-2 keV), below the direct source
detection limit.  Moreover, we have estimated the fraction of the
total number of sources represented by any population of sources
harder than the AGNs.

In section 2 we give a brief description of the data used in the
analysis. In section 3 we explain how we used the P(D) analysis
technique to derive the spectral shape of the sources dominating the
fluctuations. Finally in Section 4 we discuss the implications of our
study and the fraction of sources harder than the AGNs compatible with
fluctuations results.

\section{ DATA}

The $ROSAT$ PSPC data used in the analysis are listed in Table 1. Some
of the images were extracted from the UK Deep Survey
(Branduardi-Raymont et al. 1994) and are labelled by UKDS in the
table. The other ones - those beginning with LHDS - come from the
observations of the Lockman Hole (Hasinger et al. 1993). These two
surveys were selected due to their low galactic absorbing column and
their high exposure time. At such a low galactic H{\small I} absorbing columns,
observations above 0.5 keV are insensitive to the precise value of the
H{\small I} column density. The event files were extracted from the LEDAS
archive and were reduced using the standard package {\small ASTERIX}.

Both the original exposure time and the clean time are listed for each
image. Clean time refers to the remaining periods after suppressing
the intervals where the mean PSPC count rate was excessively high
(contamination by geocoronal or auroral backgrounds, solar scattering,
unfavourable viewing conditions). We have also rejected data intervals
where severe particle background contamination could not be modelled
accurately (Master Veto Rate $>$ 170) (see Snowden et al. 1994 and
references therein).

The R4, R5, R6 and R7 bands (as defined by Snowden et al. 1994)
correspond to energies (10 per cent of peak response) $\sim 0.44$ to $\sim
1.01$ keV, $\sim 0.56$ to $\sim 1.21$~keV, $\sim 0.73$ to $\sim
1.56$~keV and to $\sim 1.05$ to $2.04$ keV respectively and
consequently they span all of the 0.5-2 keV energy range.  Moreover,
they do not cover the softer energies (0.1 - 0.4 keV) where the
internal background as well as the contamination produced by the
Galactic component (both in emission and absorption) of the diffuse
background are more important.  We also analysed the images called
HARD in the table to compare our results with those obtained by
previous works (Barcons et al. 1994, Hasinger et al. 1993).

\begin{figure*}
  \vbox to 0cm{\vfil}
\epsfverbosetrue
\epsfysize= 200pt
\epsffile{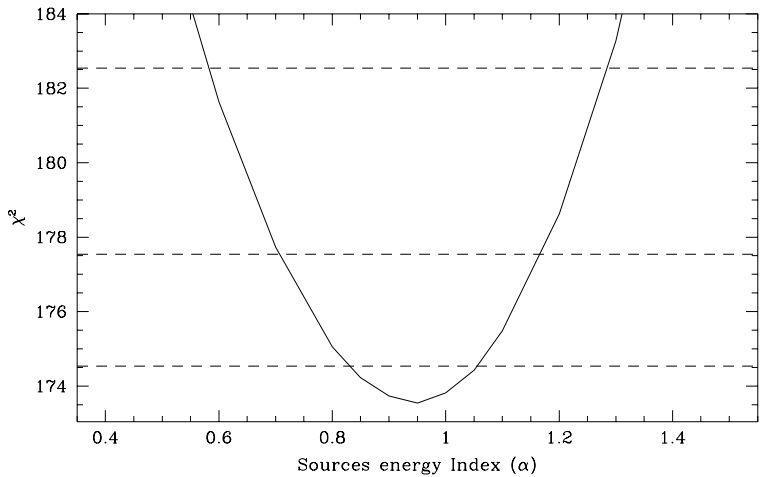}
\caption{$\chi^2$ as a function of the index $\alpha$ taking $K$ and
$\gamma$ as uninteresting parameters. Also represented as horizontal
lines are the 1$\sigma$, 2$\sigma$ and 3$\sigma$ confidence
levels (from bottom to top).}
\label{fig:chi_idx}
\end{figure*}

Only the inner part (squares of 30$\times$30 arcmin) of the PSPC field
of view was used for our analysis in order to reduce systematic
effects and uncertainties such as vignetting and variations of the
Point Spread Function (PSF). The PSF was modelled as a two dimensional
Gaussian with dispersion 10 arcsec which corresponds to the measured
PSF for on-axis sources (Hasinger et al. 1992). Finally the images
were extracted in 1 arcmin$^2$ pixels so that the specific shape of
the PSF is almost irrelevant and to avoid the pixel intensities being
correlated. That ensures that our statistical analysis of the
distribution of the fluctuations, based on the assumption that each
intensity is an independent measurement, is a good approximation.

\section{ANALYSIS}

First used in radio astronomy (Scheuer 1957, Condon 1974) the
fluctuations analysis technique (also called the P(D) analysis) has
been lately used in X--ray astronomy to obtain the number of sources
per flux interval (the $n(S)$ curve) below the direct source detection
threshold (Shafer 1983 for {\it HEAO-1} A2 data, Barcons \& Fabian
1990, for {\it Einstein Observatory} IPC images; Barcons et al. 1994
for {\it ROSAT} images; Butcher et al. 1996 for $Ginga$ pointings).

The underlying idea is that the shape of the distribution of
intensities is sensitive to fluxes down where there is about one
source per beam. Extrapolating previous fits to the $n(S)$ curve
(Branduardi-Raymont et al. 1994, Hasinger et al. 1993) it is found that
the lowest flux where this technique is applicable (one source per
beam level) is $\sim 10^{-16} \rm \ergpcmsqps$. However, one has to
take into account that the noise in the $ROSAT$ PSPC observations,
even with a pixel of $1\, {\rm arcmin}^2$, is dominated by Poisson
photon counting noise. It then happens that this noise is equivalent
to a flux $\sim 10^{-15}\, \rm \ergpcmsqps$ much larger than
the one source per beam level.  In this paper we adopt the
conservative view (as in Barcons et al. 1994) that the fluctuation
analysis can hardly say anything below fluxes equivalent to the photon
counting noise.  Therefore we expect our analysis to be sensitive from
$1-2\times 10^{-15}\, \rm \ergpcmsqps$ up to $5-6\times 10^{-15}\, \rm
\ergpcmsqps$ which is approximately the flux where the intensity
histograms rise significantly, proceeding from the high-intensity
tail downwards. Pixels with intensities larger than these fluxes will
typically contribute only  the highest intensity bin in the P(D) curve which
we have selected to have 15 pixels (see below).

The observed intensity distribution is fitted with a theoretical model
which basically includes the PSF, the differential source counts
function $n(S)$ and the photon-counting noise. Detailed descriptions
of this technique can be found in Scheuer (1974), Fabian
(1975), Shafer (1983), Warwick \& Stewart
(1989), Barcons (1992).

We have carried out a fluctuation analysis in the four bands mentioned
above (R4, R5, R6 and R7). In addition to obtaining the source counts
at fluxes below the detection threshold, this four-band analysis has
allowed us to constrain the average spectrum of the sources existing
at these low fluxes. With only one spectral band the change of the
spectral index of the sources just results in a rescaling of the
normalization factor in the $n(S)$ curve.

We parametrised the $n(S)$ curve (in the 0.5-2~keV band) as a broken
power law with an Euclidean slope above the break:

\begin{equation}
n(S)=\left\{ \begin{array}{ll}
\frac{K}{S_B}\,\left(\frac{S}{S_B}\right)^{-\gamma} & \mbox{$S<S_B$}\nonumber\\
\frac{K}{S_B}\,\left(\frac{S}{S_B}\right)^{-2.5} & \mbox{$S>S_B$}
	     \end{array}
     \right.
\end{equation}

\noindent where the fluxes also refer to the 0.5-2~keV band.  The break flux
($S_B$) is taken to be $2.2\times 10^{-14}\,\ergpcmsqps$ (Hasinger et al. 1993,
Branduardi-Raymont et al. 1994) although its exact value is not very important here due to
the fact that we are working at fluxes well below this flux
break. Both the normalization and the slope below the break ($\gamma$)
will be obtained from the fitting to the intensity distribution
curve. Assuming that the spectra of all the sources are described with
a power law model with the same `average' energy index $\alpha$
(i.e. there is no segregation in flux), the fluxes $S$ and $S_B$ can
be translated to each band taking this index into account. The
normalization $K$ of the curve $n(S)$ is then the same in all the
spectral bands.

Once the intensity distribution from the source counts is derived we
add a constant ($P_m$) to all intensities in such a way that the mean
intensity matches the measured one ($I_m$). Finally the distribution
in counts per pixel is derived by considering Poisson photon-counting
noise on this distribution.

However due to the limited amount of pixels and the distribution of
intensities, the mean total count number per pixel ($I_m$) is poorly
determined (see Table 1). To take into account this effect we let the
Poissonian mean vary and add a new term to the $\chi^2$ obtained from
the parameters of the sources. This is considered as a new
uninteresting parameter to be fitted by the model. Then we minimize
$\chi^2$ with respect to $P_m$:

\begin{eqnarray}
\chi^2(K,\gamma,\alpha)& = & {\rm MIN}_{P_m}\left[\chi^2(K,\gamma,\alpha,P_m)\right.\nonumber\\
                       &   & \left. + \left(\frac{I_m- P_m - S_m}{\sigma}\right)^2\right] 
\end{eqnarray}

\noindent where $\sigma$ is the standard error in the measurement of
$I_m$, $S_m$ is the average intensity contributed by the sources  and
$\alpha$ is the spectral energy index of the sources.

\begin{figure*}
  \vbox to 0cm{\vfil}
\epsfverbosetrue
\epsfysize= 200pt
\epsffile{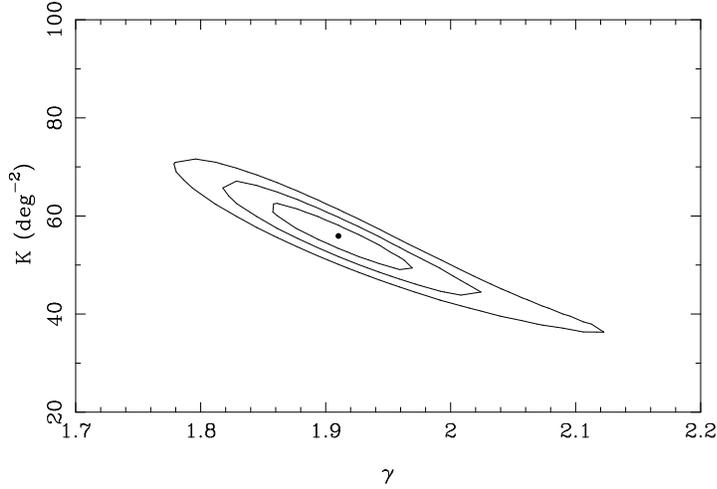}
\caption{Confidence contours for $1\sigma$ (innermost curve), $2\sigma$, 
and $3\sigma$ (outermost curve) levels in the ($K,\,\gamma$) parameter
space for the simultaneous fit to the UKDS and LHDS Deep Surveys and
$\alpha$=0.95. The point represents the best fit.}
\label{fig:conts}
\end{figure*}

In this way those sets of parameters ($K,\,\gamma,\,\alpha,\,P_m$)
with a value of $P_m$ quite far from $I_m-S_m$ give rise to high
values of the additional term in $\chi^2$ and therefore are less
likely.  To compare this approach with previous work, we note that
Barcons et al. (1994) allowed the average count per pixel to
vary freely between the measured $\pm 2\sigma$ interval, but without
adding the extra term in the $\chi^2$. A similar computation was
performed by Hasinger et al. (1993) allowing for a 5 per cent
variation in the mean. We believe that the approach we use here is
more robust, since it takes into account the measured value of the
mean with its uncertainties as an extra observational parameter to be
fitted by the model.

The histograms obtained are rebinned to have at least 15 pixels
falling in each bin since otherwise $\chi^2$ statistics would
hardly be applicable. This is due to the fact that we estimate the
dispersion in the distribution of pixel intensities based on gaussian
statistics on the number of pixels per bin which therefore has to be
large enough. Then a grid of $\chi^2$ is produced (one value for each
set of three parameters $K,\,\gamma,\,\alpha$) to find the
best-fitting parameters of the $n(S)$ curve and also the confidence
contours.

We first applied this analysis to the HARD bands (channels 40-200)
assuming $\alpha=1.0$ in order to compare it with previous works. We
found the minimum $\chi^2$ in $\gamma=1.88^{+0.28}_{-0.15}\, (2\sigma$
errors) for the UKDS-HARD band and in $\gamma=2.09_{-0.25}^{+0.08},
\,(2\sigma$ errors) for the LHDS-HARD band which closely reproduce
the best fits found by Barcons et al. (1994) and Hasinger et
al. (1993) respectively.

Then we applied the same process to each of our eight images (UKDS-R4,
UKDS-R5, UKDS-R6, UKDS-R7, LHDS-R4, LHDS-R5, LHDS-R6 and LHDS-R7) for
several values of $\alpha$ ranging from $\alpha=0.4$ to $\alpha= 1.5$
with the aim of measuring the average spectrum of the sources
dominating the fluctuations. Since we want to fit all the images
simultaneously we minimized the sum of the $\chi^2$ in $K$ and
$\gamma$, for each $\alpha$:

\begin{eqnarray*}
\chi^2_{tot}(K,\gamma,\alpha)=\chi^2_{\rm UKDS-R4}(K,\gamma,\alpha)+ \chi^2_{\rm UKDS-R5}(K,\gamma,\alpha)\nonumber\\
 \>\>\>+ \chi^2_{\rm UKDS-R6}(K,\gamma,\alpha)+\chi^2_{\rm UKDS-R7}(K,\gamma,\alpha)\nonumber\\
 \>\>\>+ \chi^2_{\rm LHDS-R4}(K,\gamma,\alpha) + \chi^2_{\rm LHDS-R5}(K,\gamma,\alpha)
\end{eqnarray*}
\begin{equation}
\hskip 2truecm + \chi^2_{\rm LHDS-R6}(K,\gamma,\alpha) + \chi^2_{\rm LHDS-R7}(K,\gamma,\alpha)
\label{eq:chi2tot}
\end{equation}
\bigskip
\begin{equation}
\chi^2(\alpha) = {\rm MIN}_{K,\gamma}\,\left[\chi^2_{tot}(K,\gamma,\alpha)\right]
\end{equation}

\begin{figure*}
  \vbox to 0cm{\vfil}
\epsfverbosetrue
\epsfysize= 200pt
\epsffile{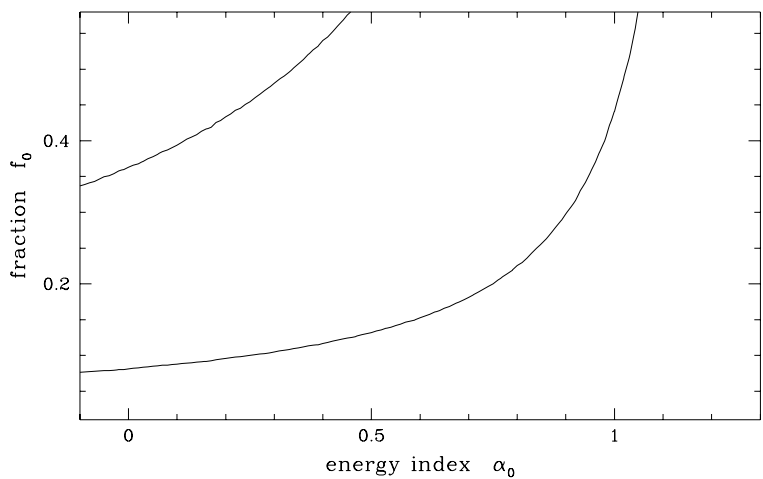}
\caption{Permitted region for a hypothetical harder population in the 
$f_0$, $\alpha_0$ parameter space (see text for details). The lines
encompass the $1\sigma$ confidence interval.}
\label{fig:f0}
\end{figure*}

The variation of $\chi^2$ for the joint fit with UKDS and LHDS Deep
surveys as a function of the spectral energy index is shown in
Fig.~\ref{fig:chi_idx}. The minimum is reached at $\alpha =
0.95^{+0.10}_{-0.15}\, (1\sigma$ errors) with a minimum
$\chi^2=173.54$ for 112 degrees of freedom. In this case
$K=55.92^{+4.13}_{-4.30}\rm deg^{-2}$ and
$\gamma=1.91^{+0.03}_{-0.04}$ which is consistent at $1\sigma$
significance level with previous fits in the HARD (0.5-2~keV) band.

Fig.~\ref{fig:conts} shows the $(K,\,\gamma)$ parameter space for this
simultaneous fit to the UKDS and LHDS images with $\alpha=0.95$ (best
fit). The point is the best fit and the contours correspond to
$1\sigma$, $2\sigma$, and $3\sigma$ significance levels. 

If fitted separately UKDS and LHDS images give different best-fits for
the sources energy spectral index and for the parameters of the $n(S)$
curve. While the fit to UKDS images shows a tendency towards high
spectral energy indices ($\alpha_{best\,fit}
\,(UKDS)\sim 1.1$) the best-fitting point for the four bands of the
LHDS image is $\alpha_{best\,fit}\,(LHDS)\sim 0.8$. Although it might
be an interesting result to establish that the slightly fainter
sources dominating the LHDS fluctuations would have harder spectra
than the UKDS ones, the difference in spectral indices between both
fits is not significant (certainly less than 2 sigma). Since, in
addition, both best fit values are also within the 1 sigma range of
the spectral energy index measured by taking both images, we interpret
this difference as being purely statistical.

\section{DISCUSSION}

We have carried out a fluctuations analysis on four different bands of
two {\it ROSAT} PSPC deep surveys simultaneously. This has enabled us
to obtain an energy index for the average spectrum of the sources
contributing to fluxes where this analysis is sensitive (0.5-2 keV
fluxes around and below $\sim 10^{-15}\, \rm \ergpcmsqps$).

The best-fitting spectral slope for the sources has been found to be
$\alpha=0.95^{+0.10}_{-0.15}$ which does not result in a very
significant flattening of the mean spectral index with regard to that
of the sources (AGN) known to dominate the source counts at higher
fluxes ($\alpha\sim 1.0-1.5$). This seems to be in contrast with the
tendency exhibited by the X-ray sources found in previous analyses of
$ROSAT$ PSPC images, according to which fainter sources have harder
spectra (Hasinger et al. 1993, Vikhlinin et al. 1995).

Since the spectral slope we derive in our analysis is an average
spectrum, we tested the possibility of the existence of two different
populations of sources contributing to the source counts at the
interval of fluxes we are dealing with (see section 2). One of these
populations would be made up by the AGNs ($\alpha\sim 1.0-1.5$, say,
$\alpha=1.2$) and the other one by sources with harder spectra. We
assigned to this last population a fraction $f_0$ of the total number
of sources at a representative flux, $S\sim 10^{-15}\,\ergpcmsqps$ and
an energy spectral index $\alpha_0$. We then calculated the total
count rate that both classes of sources would produce in each spectral
band for a given pair of values ($\alpha_0,\,f_0$). 

\begin{equation}
R_{i}^S=f_0\,(R_i)_{\alpha_0}^S\,+\,(1-f_0)\,(R_i)_{\alpha=1.2}^S
\end{equation}
with $i=4,5,6,7$ for R4, R5, R6 and R7 bands respectively.

In this equation $R_i$ gives the count rate produced by the
corresponding class of sources at flux $S$ in the $i$ band and was
calculated with the X-ray spectral analysis package {\small XSPEC} assuming
power law models. If now we assume that the whole population of
sources has the same spectrum, we can obtain this `average' energy
spectral index ($\alpha_i$) able to reproduce the count rate in each
band, $R_i$. 

As we have previously calculated the value of the $\chi^2$ resulting
from the $P(D)$ analysis for each set of parameters $\alpha$, $K$,
$\gamma$ in each spectral band (see equation~\ref{eq:chi2tot}), we can
calculate the total $\chi^2$ for each pair $\alpha_0$, $f_0$:

\begin{equation}
\chi^2(\alpha_0,f_0)=\rm MIN_{K,\gamma}\>\sum_{i,j}
\chi^2_{i,j}(\alpha_{i},K,\gamma)
\end{equation}
where $i=4,5,6,7$ as before, and $j=1$ for the UK Deep Survey images
and $j=2$ for the Lockman Hole ones.

In Fig.~\ref{fig:f0} we show the ($\alpha_0$, $f_0$) parameter space
resulting from this calculation, where the lines limit the
permitted region from the $1\sigma$ confidence interval.

The conclusion that can be extracted from this analysis is that
provided this second population exists and is made up of some type of
galaxies, for which an energy spectral index $\alpha_o=0.4-0.6$ is
being found (Romero-Colmenero et al. 1996, Carballo et al. 1995),
these sources would constitute $\sim (30\pm 20)$ per cent of the source
counts at $\sim 10^{-15}\, \rm \ergpcmsqps$.

Yet, with the data we have it is not possible to distinguish whether
both populations are segregated in flux (fainter sources with harder
spectra) or on the contrary, there is a complete mixture through all
the interval of fluxes we have been considering here.  Moreover far
from providing the solution to the X--ray background spectral paradox
this analysis has confirmed the tendency already known at higher
fluxes: the sources seem to be too soft to account for the spectrum of
the cosmic X--ray radiation at low energies.

This research has made use of data obtained from the Leicester
Database and Archive Service at the Department of Physics and Astronomy,
Leicester University, UK. Partial financial support for this work was
provided by the DGICYT under project PB92-0501.

\bsp

\end{document}